\documentclass[floatfix,%
preprint,
  prd
]{revtex4-2}

\usepackage[dvipsnames]{xcolor}
\usepackage{graphicx}
\usepackage{hyperref}
\usepackage{bm}
\usepackage{xspace}

\newcommand{\Rz}{\ensuremath{\rho^0}\xspace}
\newcommand{\Jpsi}{\ensuremath{{\rm J}/\psi}\xspace}
\newcommand{\psip}{\ensuremath{\psi(2S)}\xspace}
\newcommand{\Us}{\ensuremath{\Upsilon(1S)}\xspace}
\newcommand{\Un}{\ensuremath{\Upsilon(nS)}\xspace}
\newcommand{\Uss}{\ensuremath{\Upsilon(2S)}\xspace}

\newcommand{\Wgp}{\ensuremath{W_{\gamma \mathrm{p}}}\xspace}
\newcommand{\de}[1]{{\rm d}#1}

\begin{document}

\title{
Exclusive quarkonium photoproduction: predictions with the Balitsky-Kovchegov equation including the full impact-parameter dependence
}

\author{J. Cepila}
\affiliation{Faculty of Nuclear Sciences and Physical Engineering,
Czech Technical University in Prague, Czech Republic}
\author{J. G. Contreras}
\affiliation{Faculty of Nuclear Sciences and Physical Engineering,
Czech Technical University in Prague, Czech Republic}
\author{M. Vaculciak}
\affiliation{Faculty of Nuclear Sciences and Physical Engineering,
Czech Technical University in Prague, Czech Republic}

\date{\today}

\begin{abstract}
The ongoing Run 3 at the Large Hadron Collider (LHC) is substantially increasing the luminosity delivered to the experiments during  Run 1 and Run 2. The advent of the high-luminosity upgrade of the LHC (Run 4 to 6), as well as the improvements to all detectors, will allow for the collection of an unprecedented amount of data in the next decade. This opens the possibility of performing measurements which have been limited by the smallness of the available data samples. This is the case of multi-differential studies of $\Jpsi$, as well as of $\Upsilon$ excited states, in exclusive diffractive photon-induced interactions.
Here, we present predictions for the cross-sections of these processes utilising the dipole amplitude from the Balitsky-Kovchegov (BK) equation solved in the target rapidity and including the full impact-parameter dependence. Cross-sections are computed as a function of the photon--proton centre-of-mass energy as well as a function of Mandelstam-$t$. Ratios of cross-sections for different states and for the same state at different Mandelstam-$t$ values are also presented.  
The contribution to these observables of the non-linear terms in the BK equation is discussed.
\end{abstract}

\maketitle

\section{Introduction}
When the protons accelerated at the Large Hadron Collider (LHC) interact with an impact parameter larger than twice their radius, the process involves at least one photon, from the electromagnetic field of the incoming particles, in what is called ultra-peripheral collisions (UPCs). This area, UPCs, is currently a very active research topic at the LHC, see e.g. Refs.~\cite{Contreras:2015dqa,Klein:2019qfb,ALICE:2022wpn}.

One process that has captured the attention of both experimentalists and theorists alike is the exclusive photoproduction of vector mesons, which is attractive because of its strong sensitivity to the gluon distribution in the target particle~\cite{Ryskin:1992ui}, the possibility of studying nonperturbative (e.g. \Rz production) and perturbative regimes (e.g. charmonium or bottomonium production), and the clean experimental environment containing only a few charged particles (normally two) from the decay of the vector meson. In addition, the LHC offers a huge range of centre-of-mass energies of the photon--proton ($\Wgp$) system extending from some 20 GeV to more than one TeV~\cite{ALICE:2022wpn}.

Although the ground states of the different vector mesons have been extensively studied at the LHC, e.g. \Rz~\cite{CMS:2019awk,ALICE:2020ugp,ALICE:2024ife} or \Jpsi~\cite{LHCb:2022ahs,LHCb:2018rcm,ALICE:2023mfc,ALICE:2023jgu,CMS:2023snh}, there are no multi-differential measurements of $\Jpsi$ diffractive photoproduction yet.
There is also relatively little information about the excited states. In particular, the measurements related to the $\Upsilon$ family are few. There are some early papers from H1 and ZEUS at HERA~\cite{Breitweg:1998ki,H1:2000kis,ZEUS:2011spj}. At the LHC there is a measurement of the $\Wgp$ dependence of \Us production by the CMS collaboration~\cite{CMS:2018bbk}, and a measurement from the LHCb collaboration where the $\Wgp$ dependence of \Us states was measured along with the total cross-section, in the LHCb fiducial volume, for both \Us and \Uss~\cite{LHCb:2015wlx}. 
This situation is expected to change in the near future, thanks to the large amount of data being currently recorded by the LHC collaborations and the data to be obtained in the High-Luminosity LHC~\cite{Citron:2018lsq,Aberle:2749422}

The first computation of exclusive diffractive photoproduction of quarkonia was carried out for the case of the \Jpsi in the leading logarithm approximation of quantum chromodynamics (QCD)~\cite{Ryskin:1992ui}. Since then, this process has been studied in different frameworks and extended to include other vector mesons. Recently, motivated by the measurements performed at the LHC at energies beyond what was achieved at HERA, several groups have studied this process, a few of the most recent results are e.g. in Refs~\cite{Boer:2023mip,Goncalves:2024jlx,Flett:2024htj,Nemchik:2024lny,Peredo:2023oym,Guzey:2024gff,Cepila:2023dxn}. 
The exclusive diffractive photoproduction of $\Un$ states has been studied in the leading logarithmic approximation of QCD~\cite{Jones:2013pga,Eskola:2023oos} as well as in different implementations of the dipole picture~\cite{Kowalski:2006hc}, e.g. in Refs.~\cite{Santos:2013lpa,SampaiodosSantos:2014puz,Goncalves:2014swa,Goncalves:2016sqy,GayDucati:2016ryh,Bendova:2018bbb,Henkels:2020kju}.

Our group has studied diffractive photoproduction of quarkonia in the colour-dipole approach utilising the solutions of the Balitsky-Kovchegov (BK) equation~\cite{Balitsky:1995ub,Kovchegov:1999yj,Kovchegov:1999ua} describing the evolution in rapidity of the dipole-target amplitude~\cite{Cepila:2018faq,Bendova:2019psy,Cepila:2023pvh}. 
A key characteristic of the BK equation, discussed in Sec.~\ref{sec:bk}, is that it includes a non-linear term to account for the gluon annihilation contribution expected to be important at high energies when the regime of gluon saturation is approached (for a recent review see Ref.~\cite{Morreale:2021pnn}). The BK equation has been improved several times in the last years by including higher order corrections (e.g. Ref.~\cite{Iancu:2015joa}), formulating it in terms of the target rapidity~\cite{Ducloue:2019ezk}, and including the dependence on the impact parameter, e.g. Ref.~\cite{Berger:2012wx,Cepila:2018faq,Bendova:2019psy,Cepila:2023pvh,Mantysaari:2024zxq}.

In this article, we present predictions for the cross-section of exclusive diffractive photoproduction of charmonia and bottomonia ground and excited states as a function of $\Wgp$ and Mandelstam-$t$, the square of the momentum transferred in the interaction. We utilise the solutions of the BK equation in the target rapidity and include the full impact-parameter dependence~\cite{Cepila:2023pvh}. The predicted cross-sections are used to construct ratios of different states and of the same state at different Mandelstam-$t$ values. We also quantify the importance of the non-linear terms of the BK evolution in these observables. The rest of the article is organised as follows: in Sec.~\ref{sec:review} the formalism is reviewed including a short discussion of the BK equation in Sec.~\ref{sec:bk} and of the model used to describe the vector meson wave function in Sec.~\ref{sec:wf}; then in Sec.~\ref{sec:results} we present our predictions for cross-sections (Sec.~\ref{sec:cs}) and discuss (Sec.~\ref{sec:nl}) the contribution of the non-linear terms in the BK equation.

\section{Review of the formalism\label{sec:review}}
\subsection{Photoproduction cross-section}
The cross-section for exclusive diffractive photoproduction of a vector meson off a proton is the sum of the cross-sections for the transverse (T) and longitudinal (L) contributions given by
\begin{equation}
\frac{\de \sigma_{\rm T,L}}{\de |t|}(t, Q^2, \Wgp) =
\frac{1}{16\pi} (1 + \beta_{\rm T,L}^2) R^2_{\rm T,L} \left| \mathcal{A} _{\rm T,L} \right|^2, 
\end{equation}
where  $Q^2$ denotes the photon of virtuality. The amplitude is given by
\begin{equation}
\mathcal{A}_{\rm T,L} =
i \int \de \vec{r} \int\limits_{0}^{1} \frac{\de z}{4\pi} \int \de^2 \vec{b} \left( \Psi_V^\dagger \Psi\right)_{\rm T,L}  
 e^{-i[\vec{b}- (\frac{1}{2}-z)\vec{r}]\cdot\vec{\Delta}} 2 N(\vec{r},\vec{b}; \eta),
 \label{eq:A}
\end{equation}
where $\vec{\Delta}$ is the momentum transferred to the proton during the interaction.
This amplitude describes a  photon that  fluctuates into a quark-antiquark colour dipole travelling at an impact parameter $\vec{b}$, with the quark taking a fraction $z$ 
of the photon energy and the dipole having a size given by the magnitude of the vector $\vec{r}$ joining the positions of the quark and the antiquark in the transverse plane. 

The dipole--target amplitude $N(\vec{r},\vec{b}, \eta)$ is a solution of the  BK equation evolved in the target rapidity 
$\eta\equiv\ln(x_0/x)$ 
with $x_0=0.01$ the starting point of the evolution and $x$ given by  $x=(Q^2+M^2)/(\Wgp^2+Q^2)$, where $M$ is  the mass of the vector meson.
These solutions have been introduced and studied in detail in Ref.~\cite{Cepila:2023pvh} and are briefly presented in Sec~\ref{sec:bk}.

The wave function for the fluctuation of the photon into a quark--antiquark dipole is denoted by $\Psi$ (see e.g. Ref~\cite{Cepila:2023pvh} for its definition), while the wave function for the scattered dipole to form the vector meson is denoted by $\Psi_V$. The vector-meson wave functions are further discussed in Sec.~\ref{sec:wf}.
 
Finally, the  corrections to take into account the fact that the amplitude is not exactly imaginary and the skewedness of the process~\cite{Shuvaev:1999ce} are 
\begin{equation}
\beta_{\rm T,L} = \tan (\pi \lambda_{\rm T,L}/2), ~~~~~ R_{\rm T,L} = \frac{2^{2\lambda_{\rm T,L}+3}}{\sqrt{\pi}} \frac{\Gamma (\lambda_{\rm T,L} + 5/2)}{\Gamma (\lambda_{\rm T,L} + 4)},
\end{equation}
with 
\begin{equation}
    \lambda_{\rm T,L} = \Bigg|\frac{\partial \ln \mathcal{A}_{\rm T,L}}{\partial \ln (1/x)}\Bigg|.
    \label{eq:lambda}
\end{equation}
\subsection{The BK equation\label{sec:bk}}
\begin{figure}[t]
    \includegraphics[width=.48\textwidth]{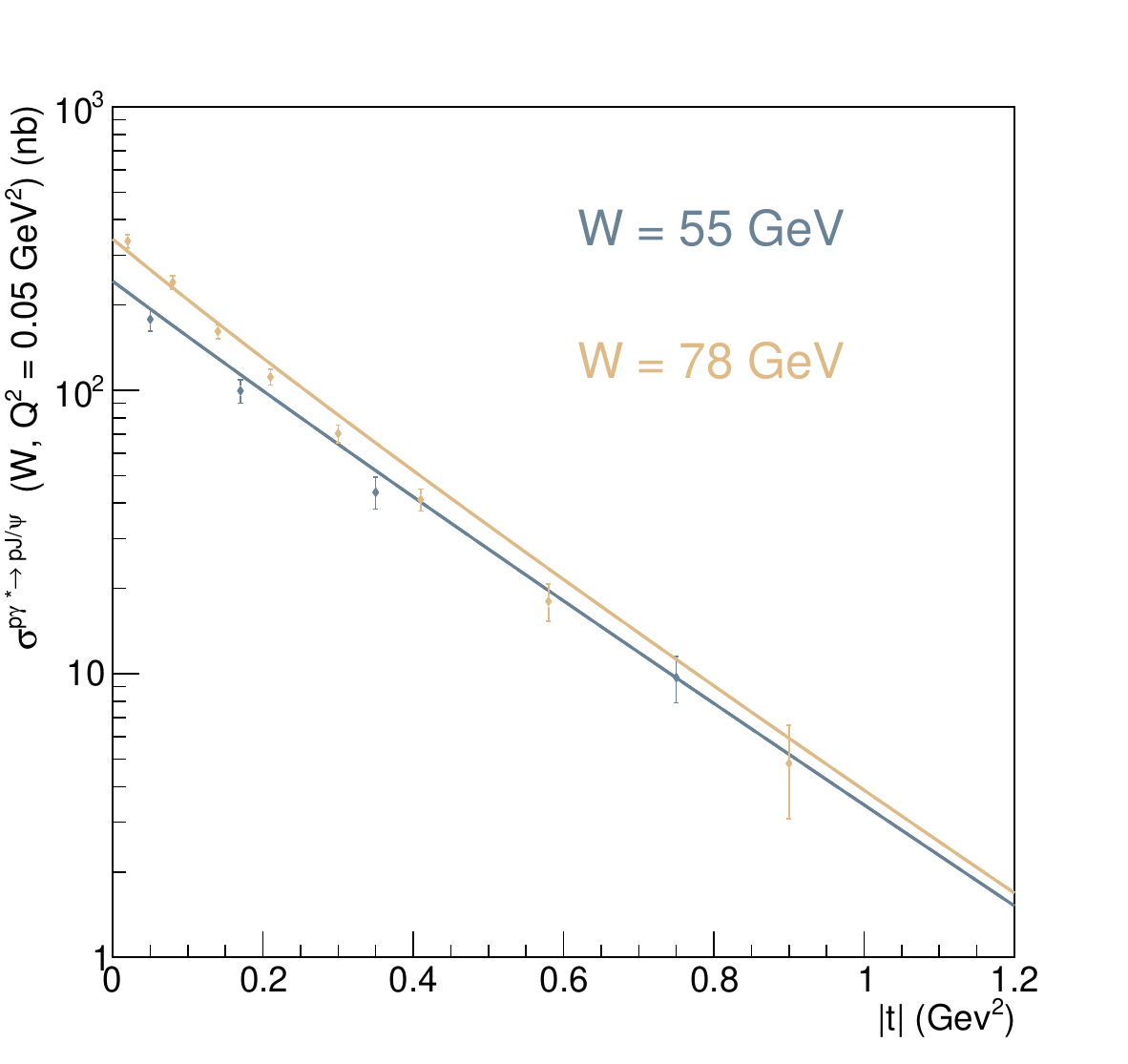}
    \includegraphics[width=.48\textwidth]{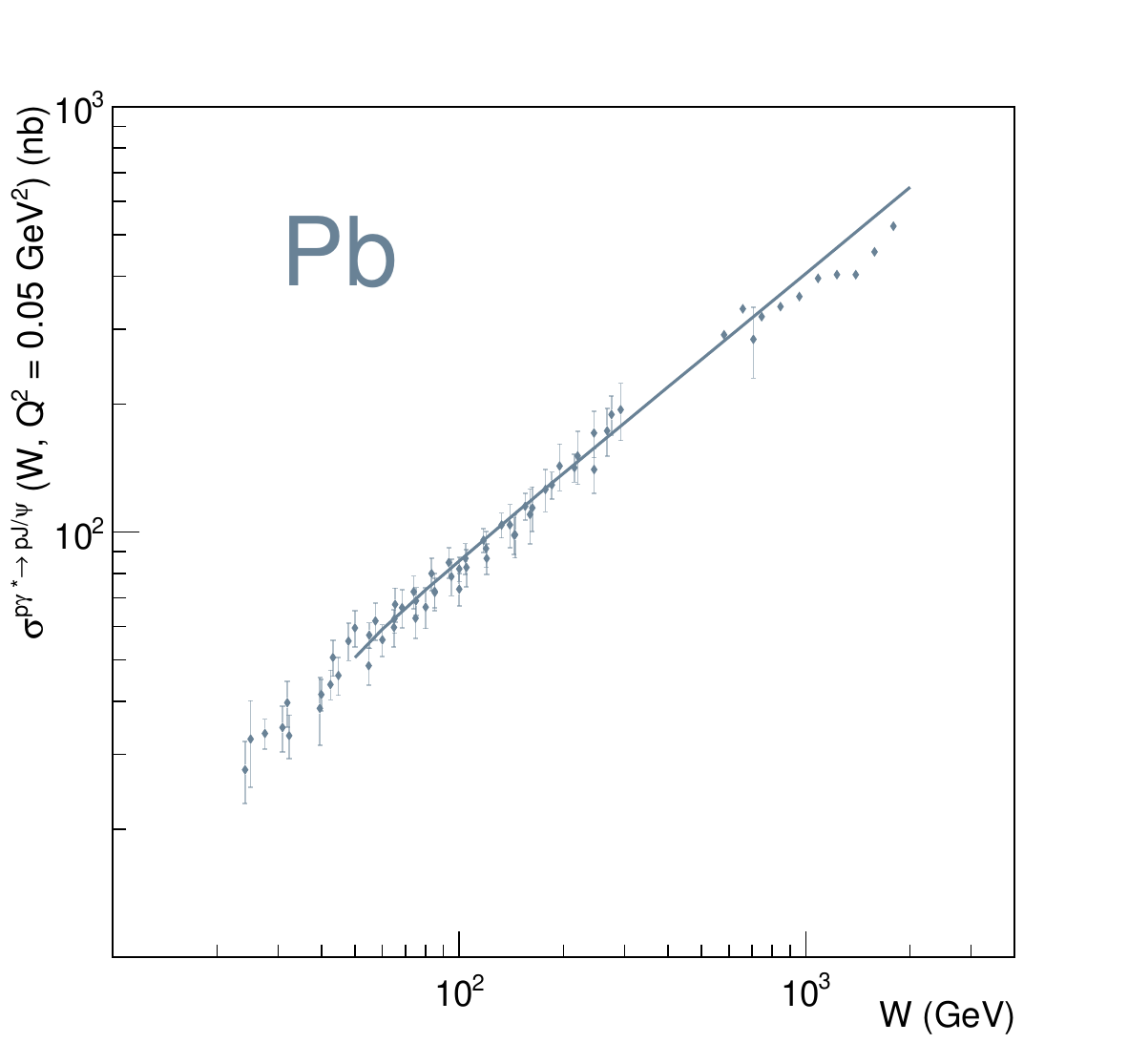}
    \caption{The $J/\psi$ production data used for extraction of the model parameters. Measurements by H1~\cite{H1:2013okq} were used for the differential cross section (left panel) and a combination of measurements by ZEUS, H1, ALICE, and LHCb~\cite{ZEUS:2002wfj, H1:2005dtp, H1:2013okq, ALICE:2014eof, LHCb:2018rcm} for the total cross section (right panel).
    \label{fig:parameter_extraction}}
\end{figure}
\begin{figure}[t]
 \includegraphics[width=0.48\textwidth]{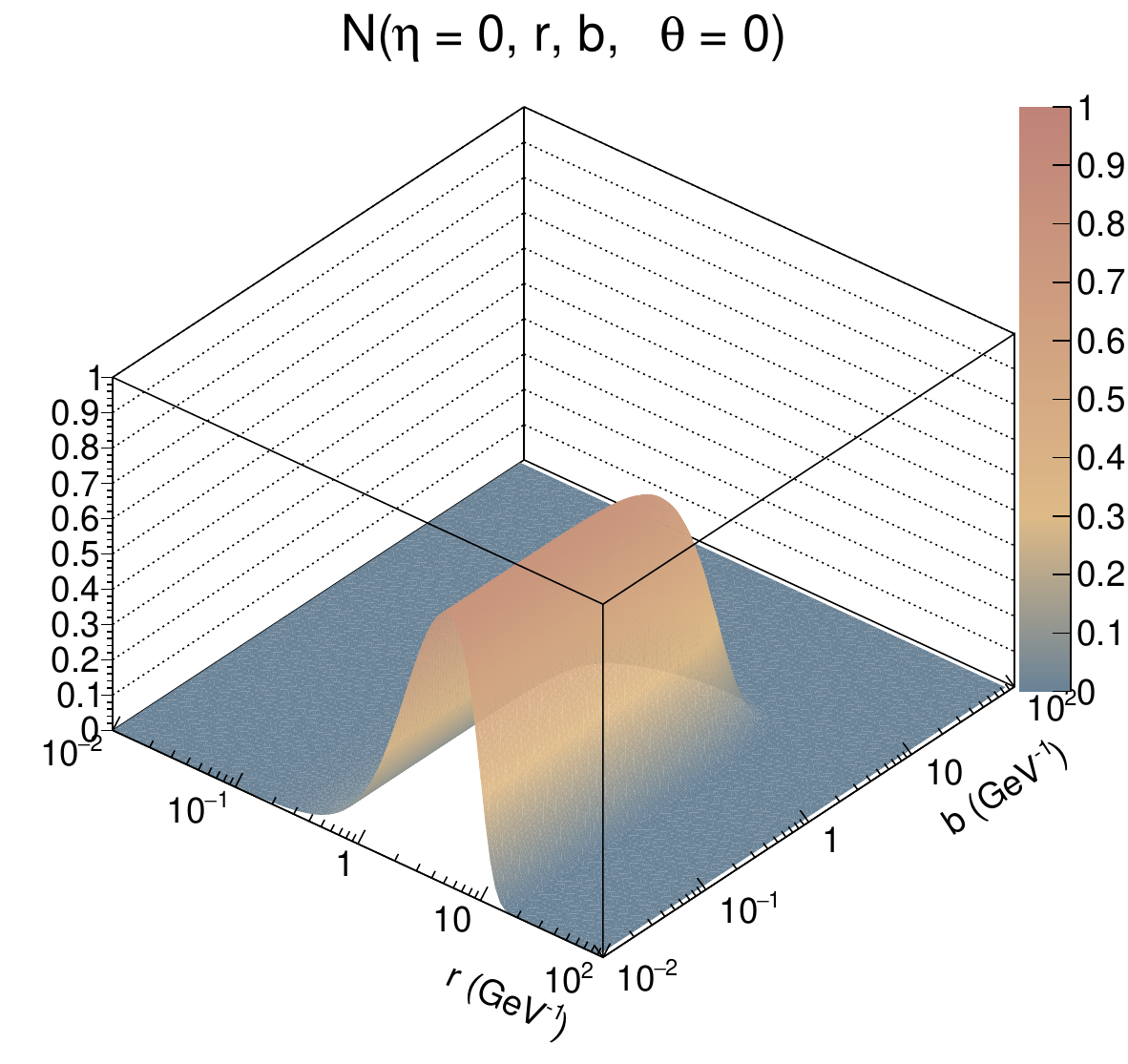}
 \includegraphics[width=0.48\textwidth]{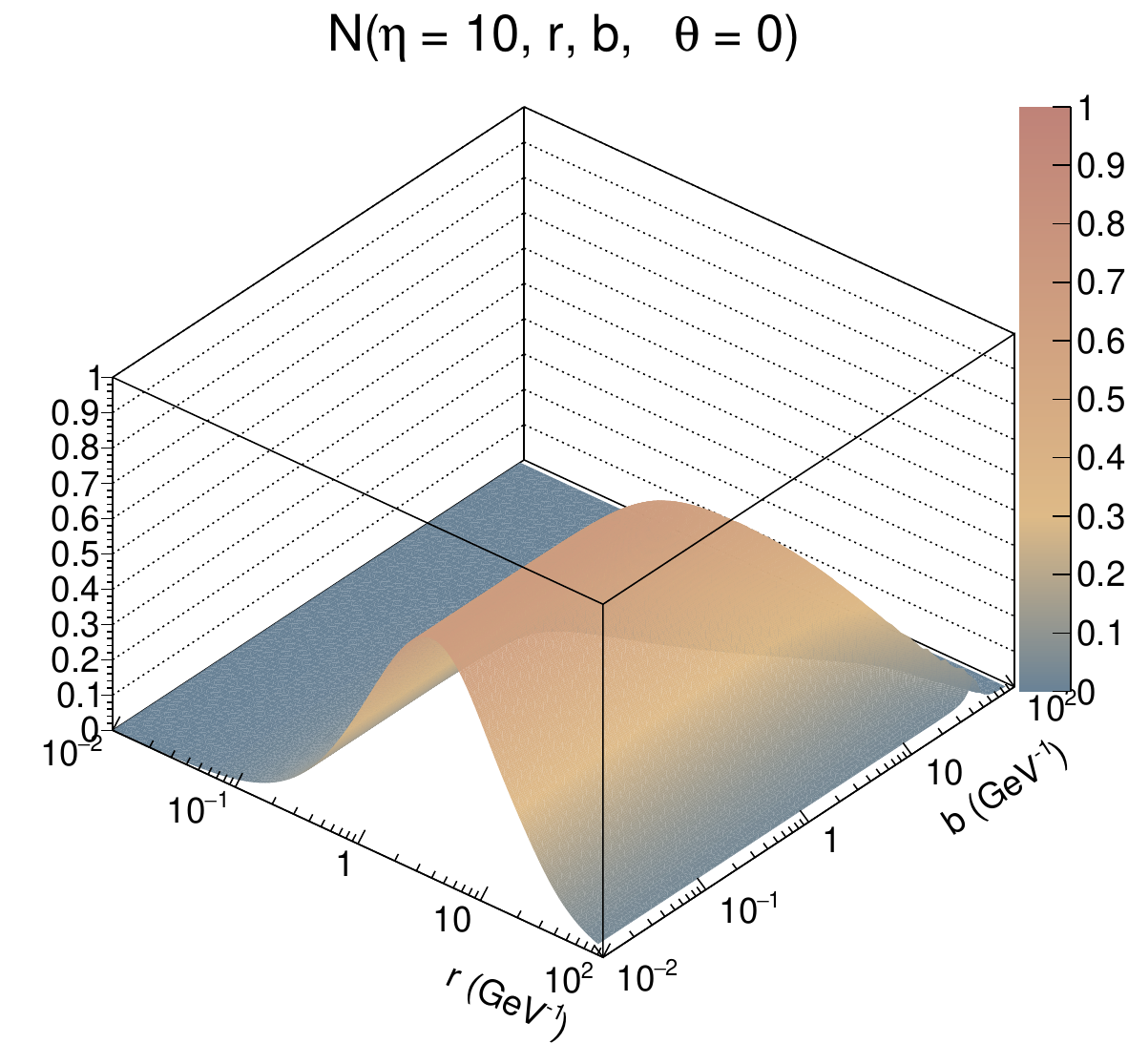}	
    \caption{Dipole amplitude at the initial condition and after evolution to rapidity 10 for the case of parallel dipole-size and impact-parameter vectors. 
    \label{fig:Nrb}}
\end{figure}

The BK equation in target rapidity was introduced in Ref. [45]. It traces the evolution with the rapidity of the dipole-target amplitude $N(\vec{r}, \vec{b}; \eta)$ when the parent dipole described by the size and impact parameter vectors $\vec{r}$ and $\vec{b}$, respectively, split into two daughter dipoles characterised by $\vec{r}_{1,2}$ and $\vec{b}_{1,2}$; the equation also includes the case when two dipoles recombine into one. The equation is given by
\begin{eqnarray}
\frac{{\rm d}N(\vec{r}, \vec{b}; \eta)}{{\rm d}\eta} = \int \mathrm{d} \vec{r}_1 K(r, r_1, r_2) \Big[& &N(\vec{r}_1, \vec{b}_1; \eta_1) + N(\vec{r}_2, \vec{b}_2; \eta_2) - N(\vec{r}, \vec{b}; \eta)	\nonumber \\
		& &- \kappa N(\vec{r}_1, \vec{b}_1; \eta_1) N(\vec{r}_2, \vec{b}_2;\eta_2) \Big], \label{eq:BK}
\end{eqnarray}
where we have introduced the constant $\kappa$ that will be used to explore the contribution of the non-linear terms by setting it to either one (the standard BK equation) or zero (the linear part of the equation)~\cite{Cepila:2020uxc}. The collinearly improved kernel is
\begin{eqnarray}
    K(r, r_1, r_2) = \frac{\bar{\alpha}_s}{2\pi} \frac{r^2}{r_1^2r_2^2} \bigg[ \frac{r^2}{\min \lbrace r_1^2, r_2^2 \rbrace} \bigg]^{\pm \bar{\alpha}_s A_1},
    \label{eq:kernel}
\end{eqnarray}
where the first part corresponds to the standard leading log approximation including the running of the strong coupling. The second, exponentiated, part comes from resumming the DGLAP-like single transverse logarithmic contributions~\cite{Ducloue:2019jmy}. 
The constant $A_1 = \frac{11}{12}$ and $J_1$ is the Bessel function. Here, $\bar{\alpha}_s = \frac{N_{\rm C}}{\pi} \alpha_s$ with \mbox{$\alpha_s = \alpha_s(\min\lbrace r, r_1, r_2 \rbrace)$} being the running strong coupling constant evaluated in the scheme with variable number of flavours.
The rapidities
$ \eta_j = \eta - \max\lbrace 0, \ln(r^2/r^2_j)\rbrace$
introduce a non-local behaviour to the equation and require a prescription to handle the rapidities earlier than the initial rapidity. Here we chose to suppress completely these contributions, that is $N(\eta<0)=0$. See Ref.~\cite{Cepila:2023pvh} for a discussion of this issue as well as for the definition of the kernel $K(r, r_1, r_2)$. The initial condition is
\begin{equation}
N(\vec{r}, \vec{b}; \eta_0)  = 1 - \exp{\left(- \frac{1}{4}(Q_{s0}^2\,r^2)^\gamma \,T(b,r)\left\{1 + c\cos(2\theta)\right\}\right)},
\label{eq:N0}
\end{equation}
where $\theta$ is the angle between the impact parameter and dipole size vectors, and the target profile is given by
\begin{equation}
T(r, b)= \exp{\left(-\frac{b^2+(r/2)^2}{2B}\right)}.
\label{eq:Trb}
\end{equation}
The values of the parameters are chosen to obtain a reasonable description of the energy evolution of $\Jpsi$ photoproduction total cross-section and the Mandelstam-$t$ dependence at $\Wgp=55$ GeV and $\Wgp=78$ GeV as shown in Fig. \ref{fig:parameter_extraction}. They are set to: $Q_{s0}^2=0.37$ GeV$^2$, $B = 3.6$ GeV$^2$, $c=0.5$, and $\gamma=1$. There is also a parameter involved in the running of the strong coupling constant which was set to $C = 2.5$. These values were updated since our previous work [42], as we extended the dataset for their extraction from solely HERA data to a combination of HERA and LHC, covering a wider energy region. Furthermore, new parameters were obtained for the vector-meson wave function using the latest PDG data, which also needed to be accounted for. The presented BK parameters were obtained using the following quark masses: $m_{\rm uds}=0.1$ GeV$/c^2$, $m_{\rm c}=1.27$ GeV$/c^2$ and $m_{\rm b}=4.18$ GeV$/c^2$. 
As an example of the evolution, Fig.~\ref{fig:Nrb} shows the dipole amplitude at the initial condition and after evolution to rapidity 10, for the case of parallel dipole-size and impact-parameter vectors. The shape remains similar throughout the evolution but the wavefronts move and the profile in impact parameter grows. Other values of $\theta$ provide a similar picture as discussed in   Ref.~\cite{Cepila:2023pvh}.

\subsection{Wave function
\label{sec:wf}}
The wave function of the vector meson contains non-perturbative QCD contributions and it has thus to be modelled. Under the assumption that the vector meson is predominantly a quark-antiquark state with a spin and polarisation structure as the photon~\cite{Nemchik:1994fp,Nemchik:1996cw} (see also Ref.~\cite{Krelina:2018hmt,Cepila:2019skb}), then
\begin{equation}
\Psi^{\ast}_{V}\Psi_{\gamma^{\ast}}\Big\vert_{\rm T}= e_f\delta_{f\bar f} e\frac{N_{\rm C}}{\pi z(1-z)}\Big(m_f^{2}K_0(\epsilon r)\Phi_T(r,z)-
(z^{2}+(1-z)^{2})\epsilon K_1(\epsilon r)\partial_r\Phi_T(r,z)  \Big)
\end{equation}
and
\begin{equation}
\Psi^{\ast}_{V}\Psi_{\gamma^{\ast}}\Big\vert_{\rm L}= e_f\delta_{f\bar f} e\frac{N_{\rm C}}{\pi}2Qz(1-z)K_0(\epsilon r)\Bigg(M\Phi_L(r,z)+
\delta\frac{m_f^{2}-\nabla_r^{2}}{Mz(1-z)}\Phi_L(r,z)\Bigg),
\end{equation}
where  $ \epsilon^{2}=z(1-z)Q^{2}+m_f^{2} $, $ N_{\rm C}=3 $ is the number of colours, $ m_f $ and $ e_f\delta_{f\bar f} $ are the fractional charge and effective mass of the quark, respectively. The Bessel functions $K_{0,1}$ arise from the wave function of the photon fluctuation into a quark-antiquark pair~\cite{Lepage:1980fj,Dosch:1996ss}. 
The scalar part $ \Phi_{\rm T,L} $ of the vector meson wave function is model-dependent. Here, we use the boosted Gaussian model~\cite{Brodsky:1980vj,Nemchik:1994fp,Nemchik:1996cw,Forshaw:2003ki} given by
\begin{equation}
 \Phi^{1S}_{\rm T,L}(r,z)=N_{\rm T,L}z(1-z)\exp\left(-\frac{m_f^{2}R^{2}}{8z(1-z)}-\frac{2z(1-z)r^{2}}{R^{2}}+\frac{m_f^{2}R^{2}}{2}\right),
 \end{equation}
\begin{equation}
\Phi^{2S}_{\rm T,L}(r,z)=\Phi^{1S}_{\rm T,L}(r,z)\left( 1+\alpha_{1}g(r,z)\right),
 \end{equation}
and
\begin{equation}
 \Phi^{3S}_{\rm T,L}(r,z)=\Phi^{1S}_{\rm T,L}(r,z)\left( 1+\alpha_{1}g(r,z)+\alpha_{2}\left(g(r,z)^2+4\left(1-\frac{4z(1-z)r^{2}}{R^{2}}\right)\right)\right),
 \end{equation}
where
\begin{equation}
g(r,z)=2+\frac{m_f^{2}R^{2}}{4z(1-z)}-\frac{4z(1-z)r^{2}}{R^{2}}-m_f^{2}R^{2}.
 \end{equation}
As it is customary, we fixed the parameters using the normalisation of the wave functions, their partial width to electron-positron pairs, and the orthogonality of the different states~\cite{Cox:2009ag,Armesto:2014sma}. 
\begin{figure}[t]
 \includegraphics[width=0.49\textwidth]{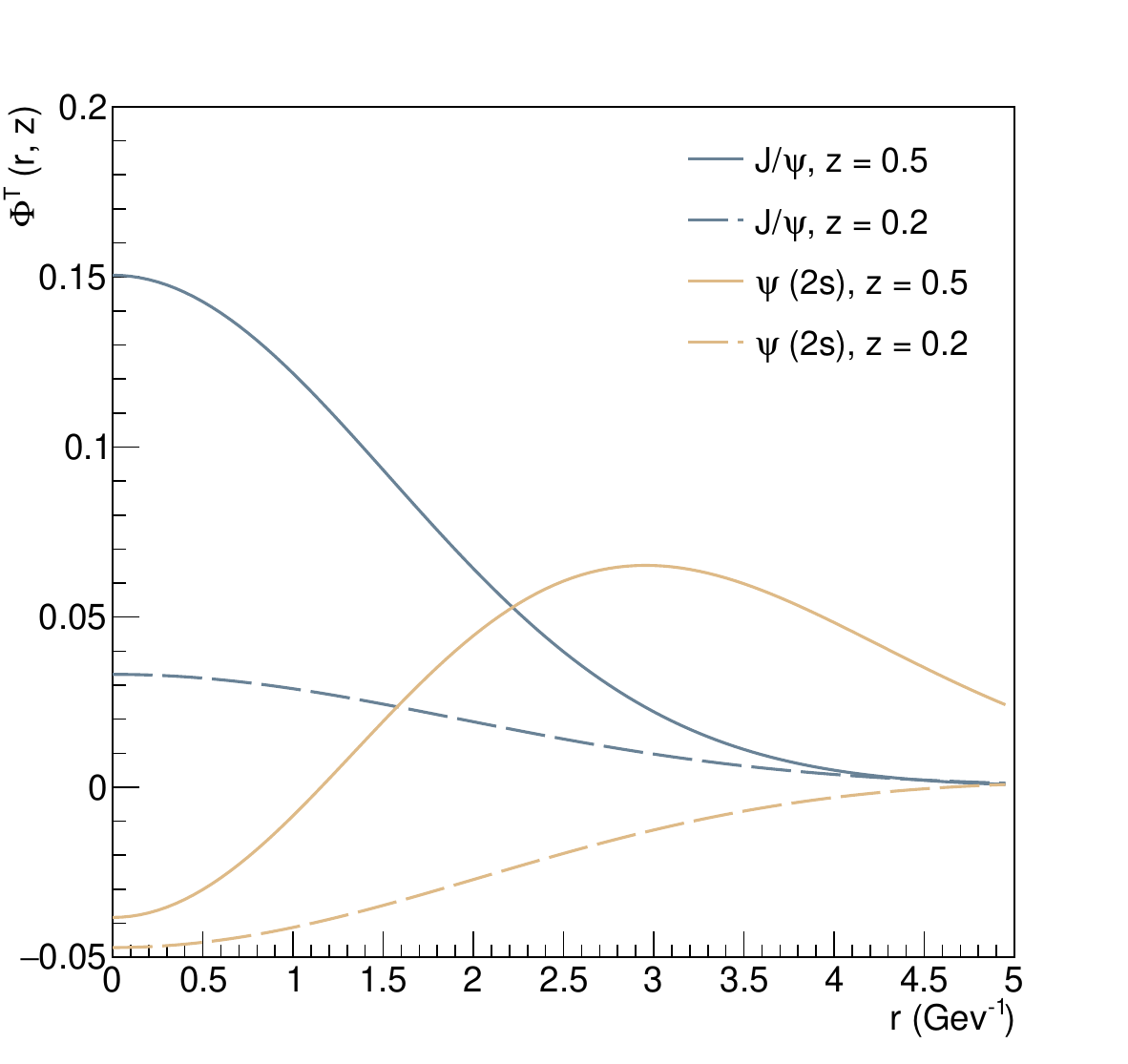}
 \includegraphics[width=0.49\textwidth]{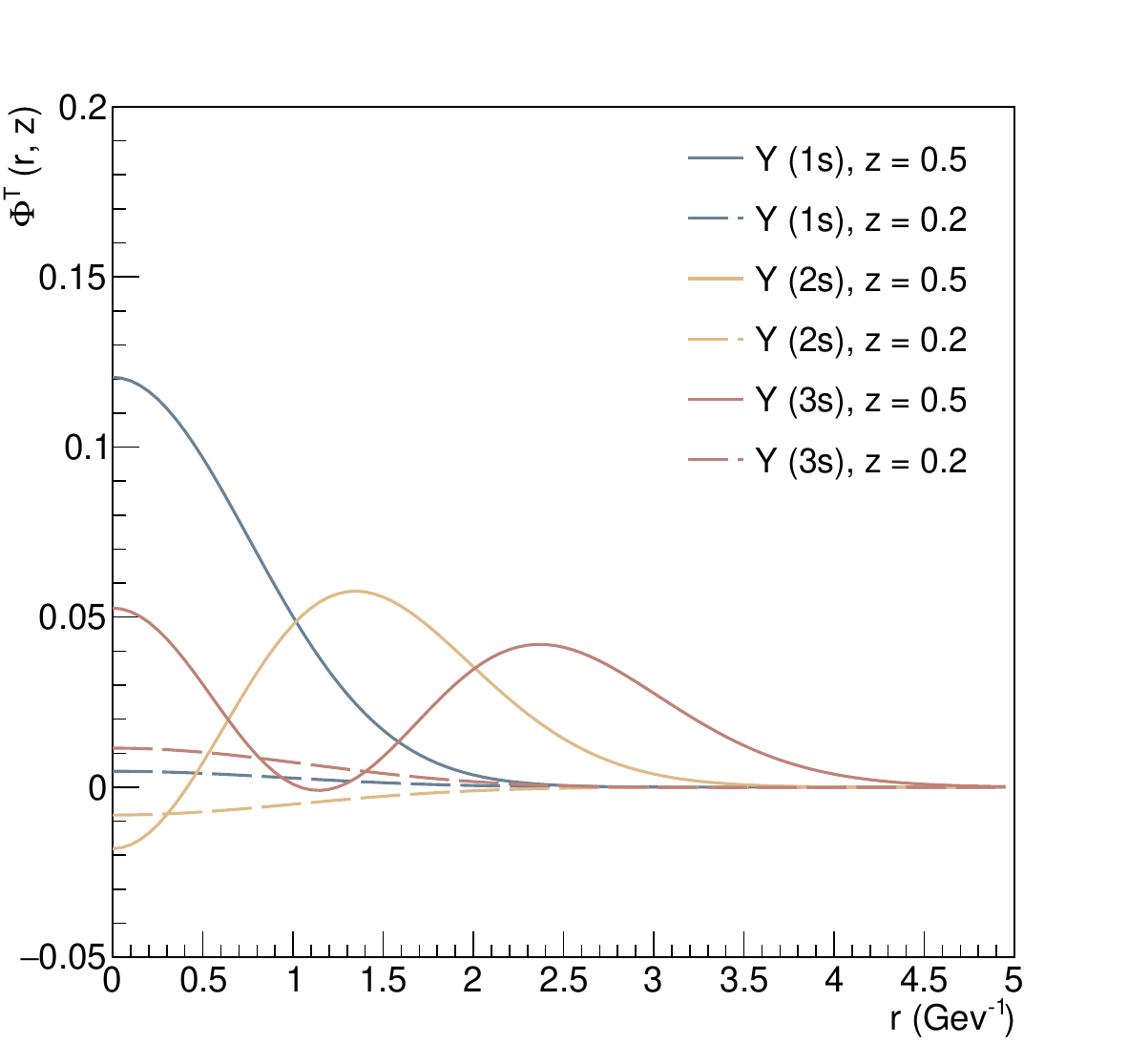}
    \caption{Scalar part of charmonium (left panel) and bottomonium (right panel) wave functions according to the boosted Gaussian model for transverse polarisation. \\
    \label{fig:wf}}
\end{figure}
The parameter values obtained by this procedure using data from PDG 2024~\cite{ParticleDataGroup:2024cfk} are reported in Table~\ref{tab:parameters}. Figure~\ref{fig:wf} shows the scalar part of the charmonium and bottomonium wave functions for the case of transverse polarisation. Note, that the case of longitudinal polarisation is almost identical to the transverse case so we do not present it here. Although the model of Gaussian distribution of the vector meson in the rest frame boosted to the proper light-cone frame is widely used and accepted ~\cite{Brodsky:1980vj,Nemchik:1994fp,Nemchik:1996cw,Forshaw:2003ki,Kowalski:2006hc,Armesto:2014sma}, there is a more sophisticated approach for modelling the distribution by solving the Schr\"odinger equation of $q\bar q$ pair with realistic potential in the rest frame of the pair~\cite{Cepila:2019skb,Krelina:2018hmt}. However, the difference between the quadratic potential (boosted Gaussian model) and more sophisticated potentials is not very big and it can be neglected in the context of these studies.  

\begin{table}[!t]
\begin{tabular}{ccccccccc}
\hline\hline
Meson & $ e_f $ & $ M_V$ (GeV) & $ m_f $ (GeV)& $ N_T $ & $ N_L $ & $ R^{2} $ (GeV$^{-2}$) & $ \alpha_{1}$  & $ \alpha_{2} $\\
\hline
$ J/\Psi $ & $ 2/3 $ & $ 3.097 $ & $ 1.27 $ & $ 0.602 $ & $ 0.598 $ & $ 2.35 $ &$0$&$0$\\
$ \Psi (2S) $ & $ 2/3 $ & $ 3.686 $ & $ 1.27 $ & $ 0.696 $ & $ 0.692 $ & $ 3.70 $ & $ -0.61 $&$0$\\
$ \phi $ & $ 1/3 $ & $ 1.019 $ & $ 0.1 $ & $ 0.958 $ & $ 0.834 $ & $ 11.1 $ &$0$&$0$\\
$ \rho $ & $ 1/\sqrt{2} $ & $ 0.775 $ & $ 0.1 $ & $ 0.930 $ & $ 0.873 $ & $ 14.2 $&$0$&$0$\\
$ \omega $ & $ 1/3\sqrt{2} $ & $ 0.782 $ & $ 0.1 $ & $ 0.927 $ & $ 0.853 $ & $ 16.6 $&$0$&$0$\\
$ \Upsilon (1S) $ & $ 1/3 $ & $ 9.460 $ & $ 4.18 $ & $ 0.482 $ & $ 0.481 $ & $ 0.57 $&$0$&$0$\\
$ \Upsilon (2S) $ & $ 1/3 $ & $ 10.023 $ & $ 4.18 $ & $ 0.623 $ & $ 0.622 $ & $ 0.82 $& $ -0.558 $&$0$\\
$ \Upsilon (3S) $ & $ 1/3 $ & $ 10.355 $ & $ 4.18 $ & $ 0.654 $ & $ 0.658 $ & $ 1.07 $& $ -1.219 $& $ 0.22 $\\
\hline\hline
\end{tabular}
\caption{Value of the parameters used in the definition of the vector meson wave functions in the boosted Gaussian model.\label{tab:parameters}
}
\end{table}

\section{Predictions\label{sec:results}}
\subsection{Cross-sections\label{sec:cs}}
\begin{figure}[t]
    \includegraphics[width=.48\textwidth]{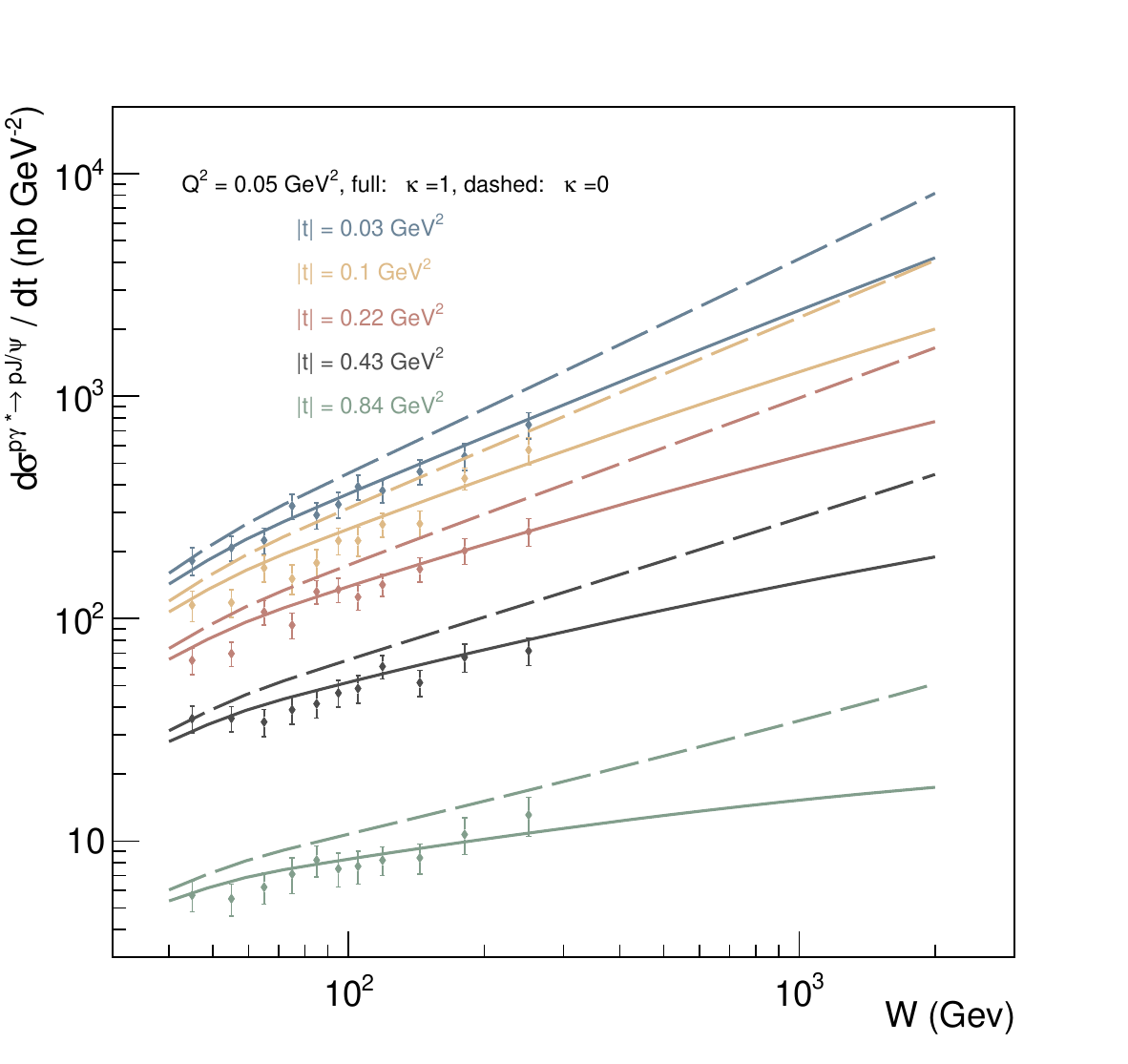}
    \includegraphics[width=.48\textwidth]{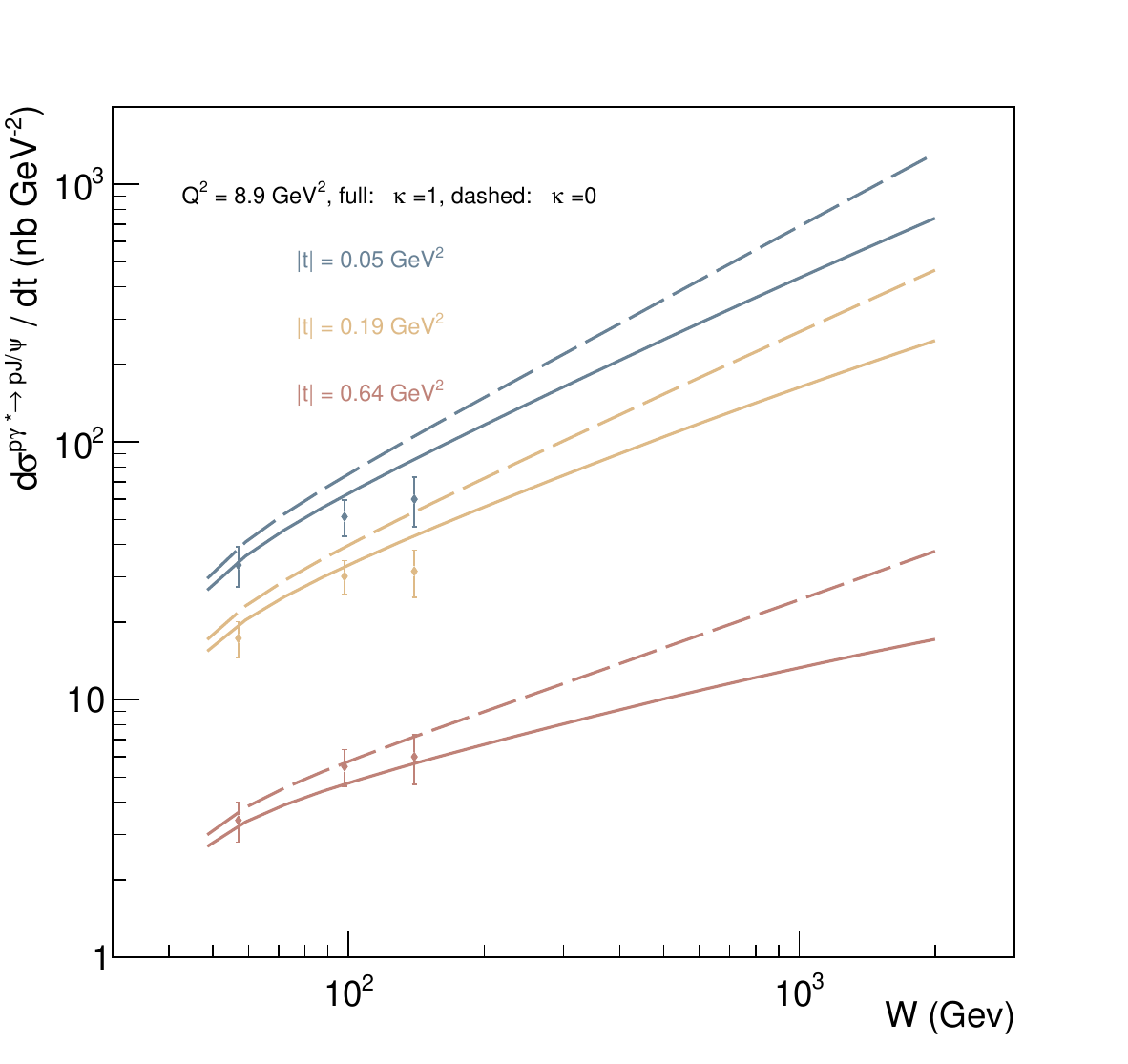}
    \\
    \includegraphics[width=.48\textwidth]{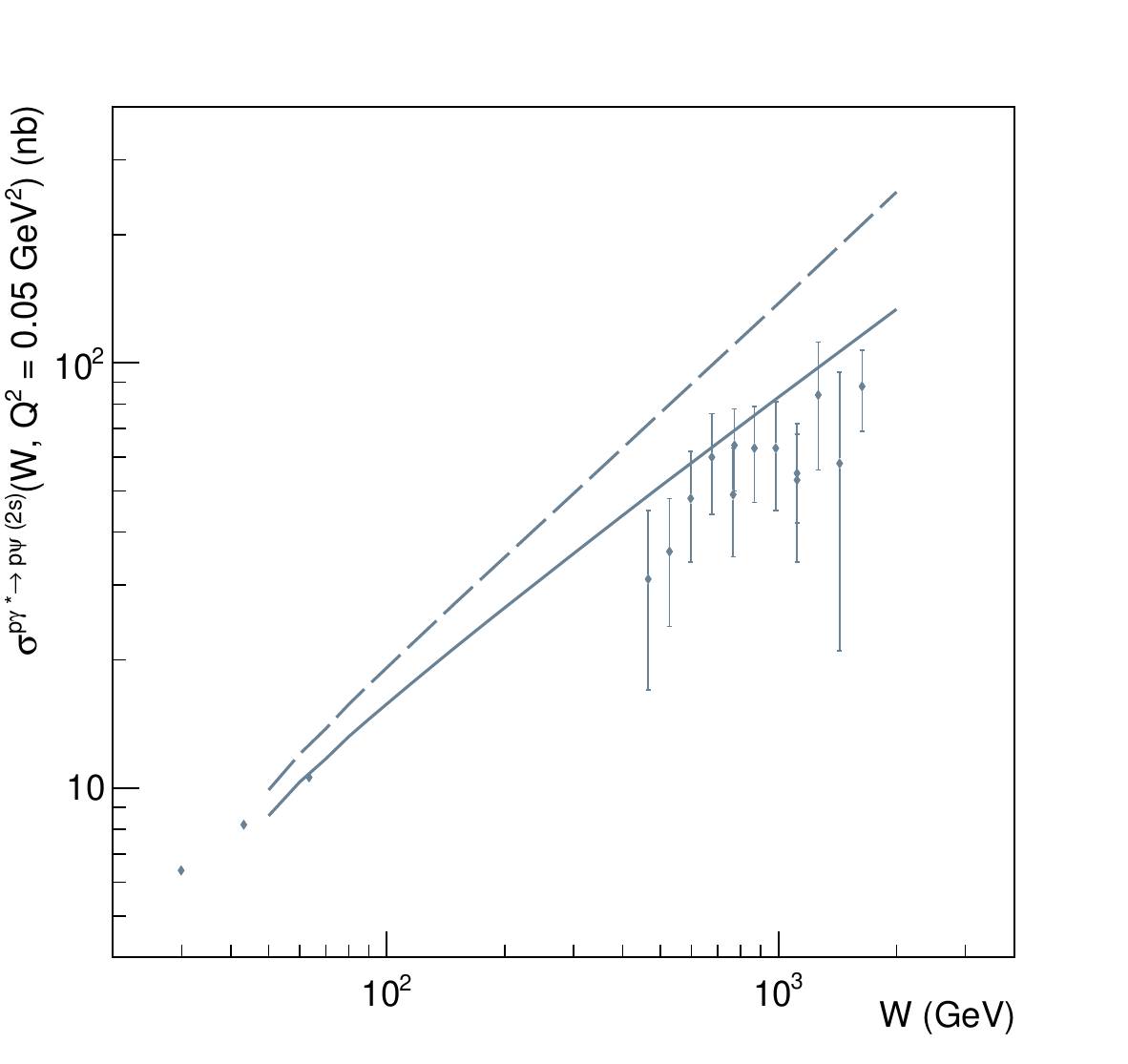}
    \includegraphics[width=.48\textwidth]{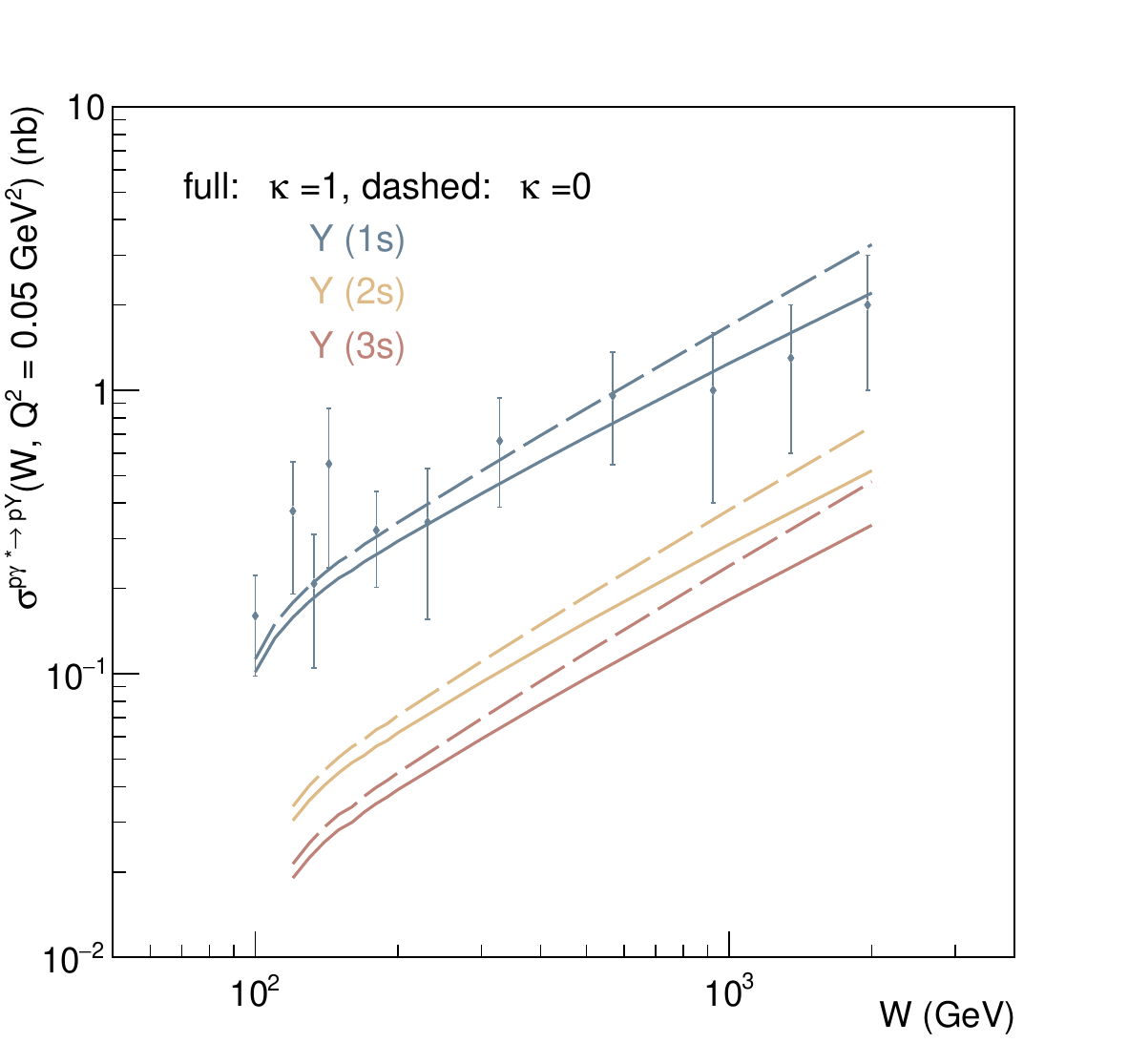}
    \caption{Energy dependence of exclusive diffractive quarkonium photoproduction. The upper panels show the comparison of the predictions with data of the H1 collaboration~\cite{H1:2005dtp} for photo- (left) and electroproduction (right). The lower panels show the prediction for the energy dependence of the $\Psi(2S)$ (left) and $\Upsilon(nS)$ (right) cross-sections integrated over Mandelstam-$t$ in the range $|t|\in (0,1.2)$ GeV$^2$. The predictions are compared to  $\Psi(2S)$ data from the LHCb collaboration~\cite{LHCb:2018rcm} and to $\Upsilon(1S)$ data from  the  H1 and ZEUS collaborations at HERA and the LHCb and CMS collaborations at the LHC~\cite{Breitweg:1998ki, H1:2000kis, Chekanov:2009zz, LHCb:2015wlx, CMS:2018bbk}
    \label{fig:qqW}}
\end{figure}
The upper panels of Fig.~\ref{fig:qqW} show the energy dependence of exclusive diffractive production of \Jpsi vector mesons for different values of Mandelstam-$t$ and photon virtualities. The predictions are compared to data from the H1 collaboration~\cite{H1:2005dtp}.
The predictions for large photon virtualities, that cannot be measured at the LHC, are of interest for the Electron--Ion Collider currently under construction~\cite{AbdulKhalek:2021gbh}.  The lower panels show the predictions for $\psip$ production (left) compared to data from the LHCb collaboration~\cite{LHCb:2018rcm} and $\Un$ production (right) compared to data from the  H1 and ZEUS collaborations at HERA and the LHCb and CMS collaborations at the LHC~\cite{Breitweg:1998ki, H1:2000kis, Chekanov:2009zz, LHCb:2015wlx, CMS:2018bbk}. The model reproduces all the measured data well.

\subsection{Contribution of the non-linear terms\label{sec:nl}}
\begin{figure}[t]
    \includegraphics[width=.48\textwidth]{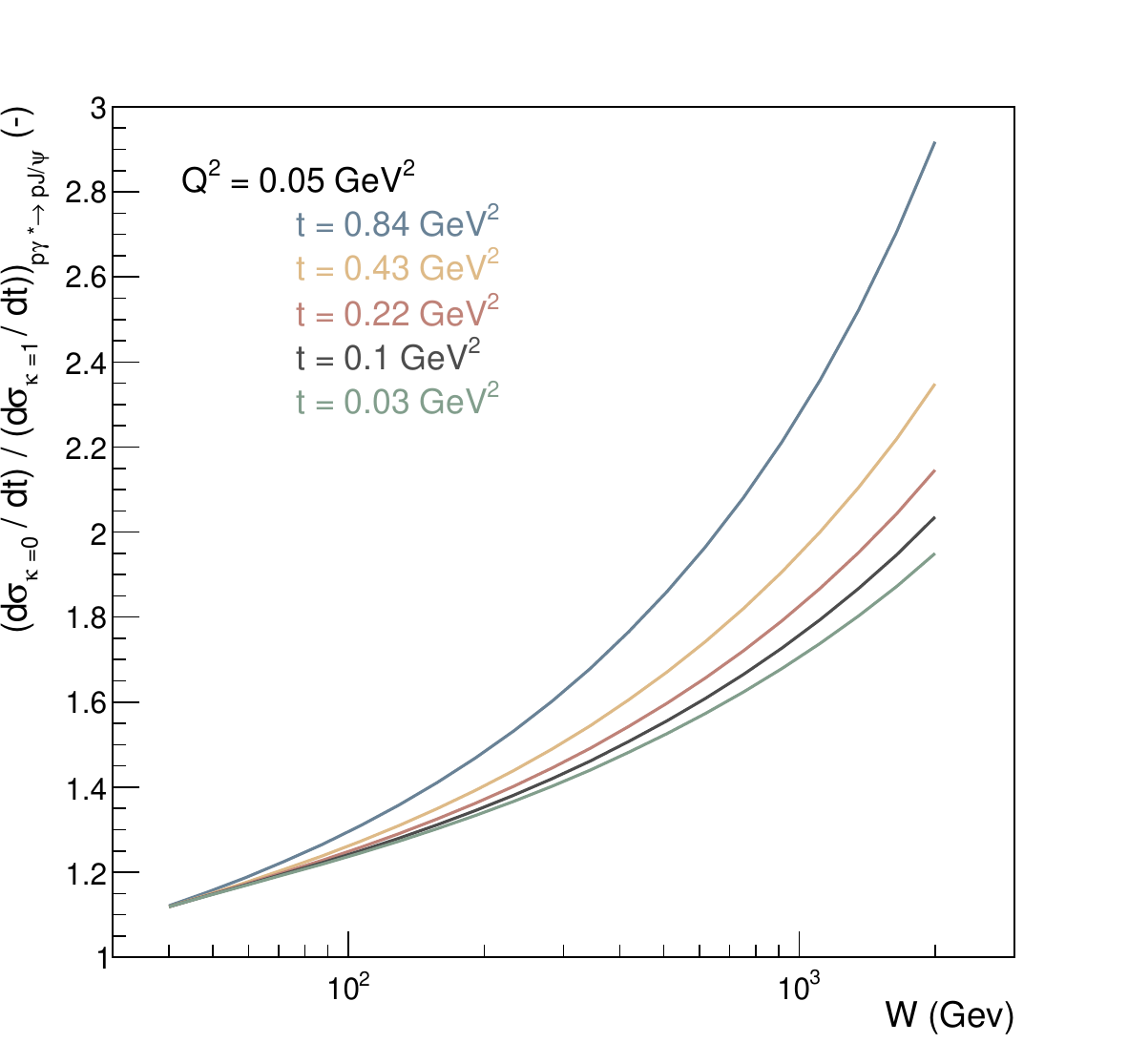}
    \includegraphics[width=.48\textwidth]{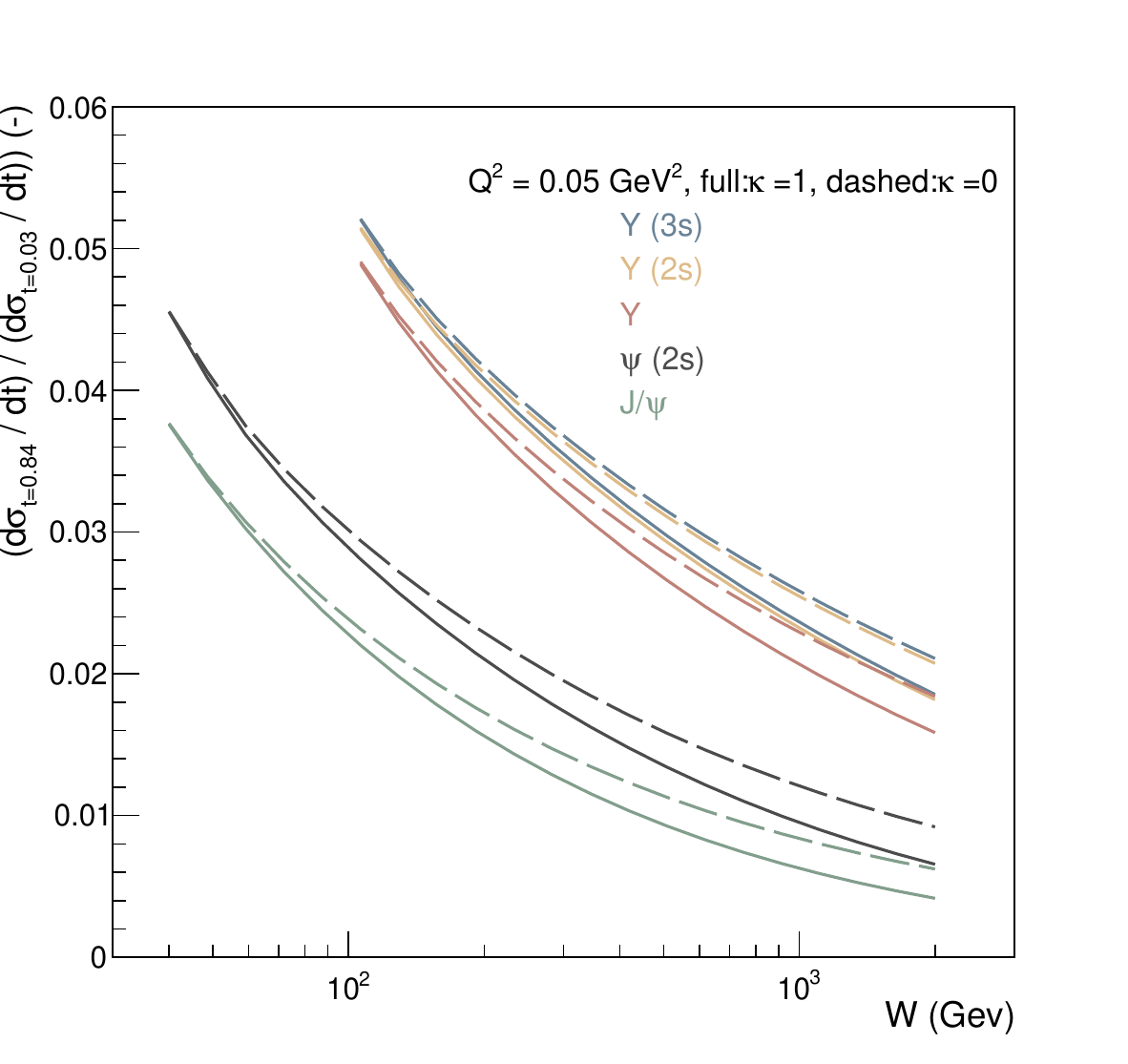}
    \caption{Left: Ratio of the cross-sections for \Jpsi exclusive diffractive photoproduction considering only the linear part of the BK evolutions, obtained by setting $\kappa=0$, to the full evolution, including non-linear terms, corresponding to use $\kappa=1$,  see Eq.~(\ref{eq:BK}).  The dependence on the value of the Mandelstam-$t$ variable is shown with the different lines. Right: Ratio of the cross-section for quarkonium exclusive diffractive photoproduction for $|t|=0.84$ GeV$^2$ to $|t|=0.03$ GeV$^2$. Solid lines represent the case $\kappa=1$ and long-dashed lines $\kappa=0$. 
    \label{fig:k}
     }
\end{figure}
\begin{figure}[t]
    \includegraphics[width=.48\textwidth]{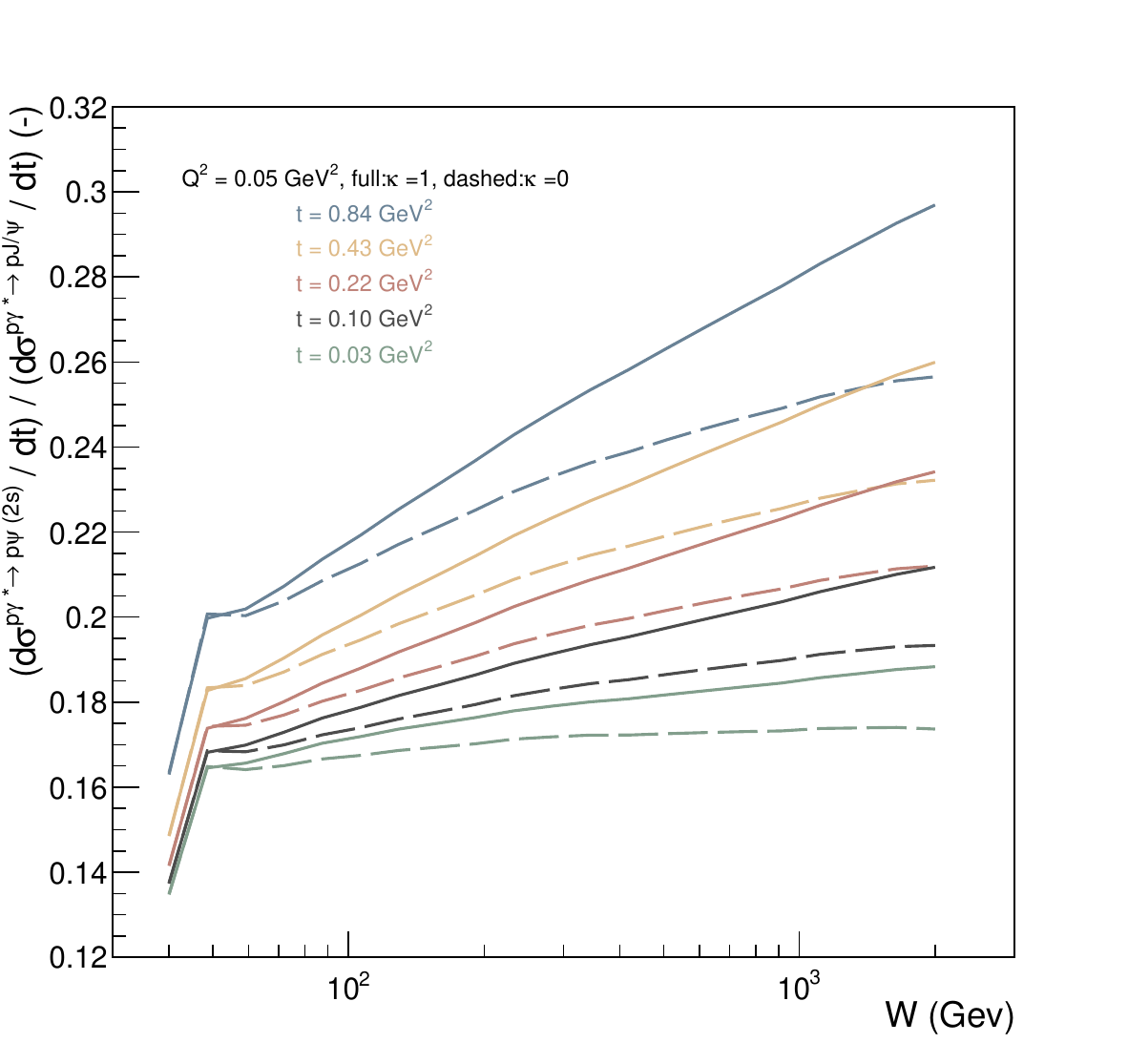}
    \includegraphics[width=.48\textwidth]{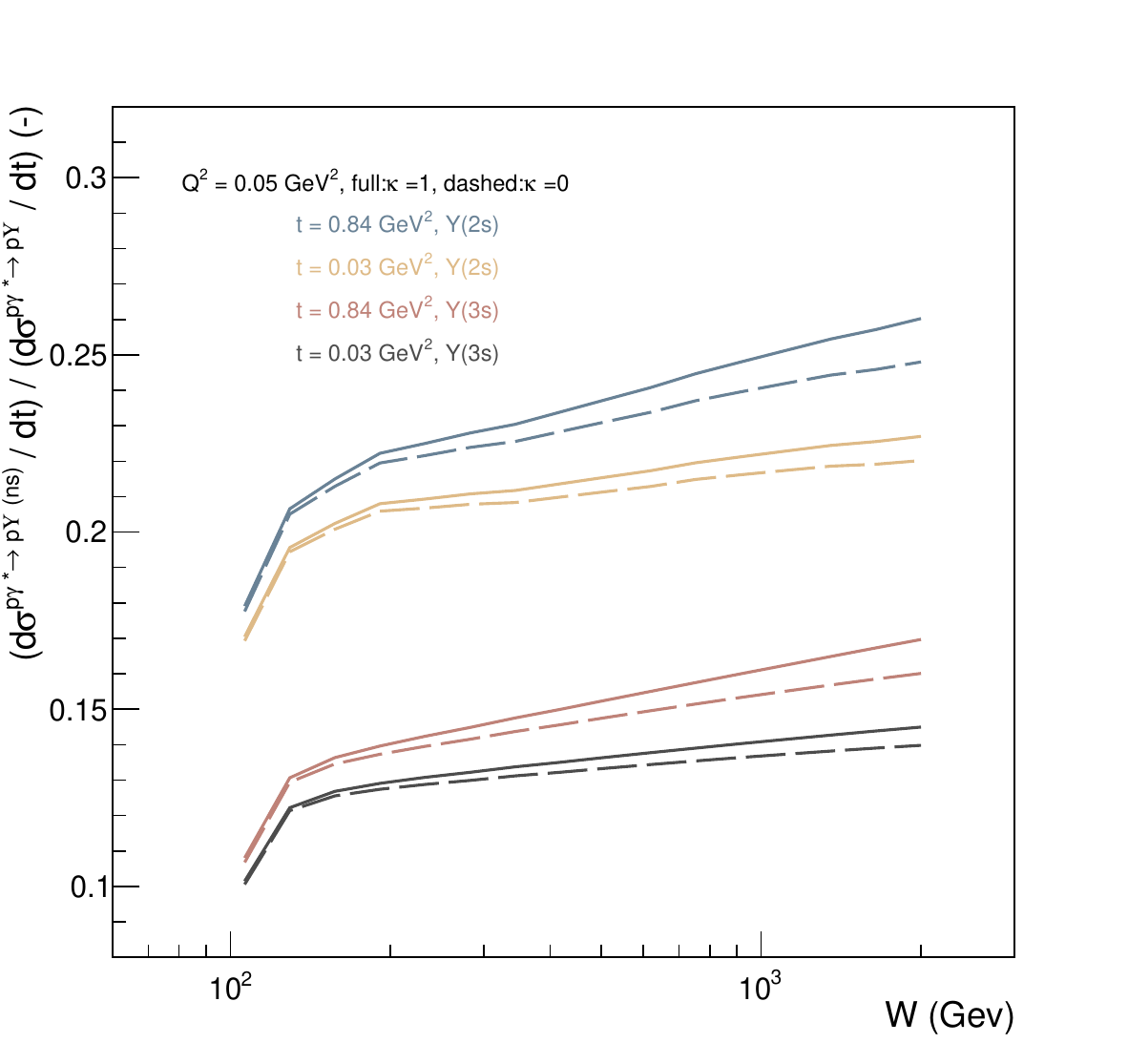} 
    \caption{Energy dependence of the ratio of the cross-sections of the excited to the ground state for the exclusive diffractive photoproduction of charmonium (left) and bottomonium (right) for different values of Mandelstam-$t$. Solid lines represent the case $\kappa=1$ and long-dashed lines $\kappa=0$.}
    \label{fig:r}
\end{figure}

To study the contribution of the non-linear terms to the cross-sections for exclusive diffractive photoproduction of quarkonium states we repeat the previous computations where we set $\kappa=0$ in Eq.~(\ref{eq:BK}) and compare with the case $\kappa=1$.
Figure~\ref{fig:k}, left panel, shows the ratio of these two cases for the production of $\Jpsi$ vector mesons. At the initial condition, there has been no evolution, thus the cross-sections for $\kappa=0$ and $\kappa=1$ are the same and the ratio is very close to one. (There is a small difference due to the effect of the corrections that involve the derivative of the amplitude, see Eq.~(\ref{eq:lambda}), which is different for the $\kappa=0$ and $\kappa=1$ cases.) As the energy increases and the evolution progresses the ratio increases signalling that the non-linear terms, that enter with a negative sign (see Eq.~(\ref{eq:BK})), slow down the evolution more and more. This pattern has a strong dependence on Mandelstam-$t$.  This can be intuitively understood from Eq.~(\ref{eq:A}), which shows that the momentum transferred is Fourier conjugate to the impact parameter, so large values of Mandelstam-$t$ sample mainly the region of small impact parameters, where the target is denser. A similar behaviour is observed for all the other vector mesons.
To explore this effect a bit further, the right panel of Fig.~\ref{fig:k} shows the energy dependence of the ratio of cross sections at large and small values of Mandelstam-$t$. There is a strong dependence of the ratio with energy, with a similar behaviour shown for all vector mesons. The effect of the non-linear corrections increases with energy but seems to be small with respect to the expected experimental uncertainties.\\
Recently, it was suggested that the ratio of the \psip to \Jpsi cross-sections for exclusive diffractive photoproduction is a good probe of the onset of saturation effects, because in the absence of non-linear effects the energy dependence is expected to be flat, while it is rising when non-linear effects become important~\cite{Hentschinski:2020yfm,Peredo:2023oym}. Figure~\ref{fig:r} shows the energy dependence of the ratio of the cross-section of excited to ground states for charmonium and bottomonium evaluated at different values of the Mandelstam-$t$ variable.
We also observe that the inclusion of non-linear effects produces a steeper rise of the ratio with energy than for the case of the linearised version of the BK equation obtained by setting $\kappa=1$. But differently than what was found in Ref.~\cite{Hentschinski:2020yfm,Peredo:2023oym}, the energy dependence of the ratio is not flat for $\kappa=0$, except for the case of charmonium production at small Mandelstam-$t$. This behaviour may complicate the use of this observable to search for the onset of saturation and calls for precise measurements in order to exploit the Mandelstam-$t$ dependence of this ratio.\\
\section{Summary and outlook}
We have presented predictions for the diffractive photoproduction of charmonia and bottomonia off protons. The predictions utilise the most recent formulation of the BK equation in the target rapidity and include the full impact-parameter dependence. The values of the parameters to describe the vector-meson wave function have been extracted using the most recent values reported in the PDG. The contribution of the non-linear terms to the cross sections, their energy evolution, and their sensitivity to different values of Mandelstam-$t$ have been explored. These predictions can be contrasted with the upcoming measurements from the LHC and the EIC facilities.

\section*{Acknowledgements}
This work was partially funded by the Czech Science Foundation (GA\v{C}R), project No. 22-27262S and 
by the FORTE project, CZ.02.01.01/00/22 008/0004632, co-funded by the European Union.

\bibliography{bibliography}

\end{document}